\documentclass{aa}
\usepackage{txfonts}
\usepackage{graphicx}

\def\msun{\hbox{M$_\odot$}}

\begin{document}

\title{Probing the farthest star clusters to the Small Magellanic Cloud}

\author{Andr\'es E. Piatti\inst{1,2,}\thanks{\email{andres.piatti@fcen.uncu.edu.ar}},
D.M.F. Illesca\inst{1,2}, M. Chiarpotti\inst{1,2}, and R. Butr\'on\inst{1}}

\institute{Instituto Interdisciplinario de Ciencias B\'asicas (ICB), CONICET-UNCuyo, Padre J. Contreras 1300, M5502JMA, Mendoza, Argentina;
\and Consejo Nacional de Investigaciones Cient\'{\i}ficas y T\'ecnicas (CONICET), Godoy Cruz 2290, C1425FQB,  Buenos Aires, Argentina\\
}

\date{Received / Accepted}

\abstract{The Small Magellanic Cloud (SMC) has been tidally shaped by the interaction
with the Large Magellanic Cloud (LMC). The scope of such an interaction has recently
been studied from different astrophysical properties of its star cluster population,
which point to star clusters placed remarkably outside the known extension of the galaxy.
In this work we report results for three of the recently identified most external SMC star 
clusters, OGLE-CL-SMC0133, OGLE-CL-SMC0237, and Lindsay~116, using deep GEMINI GMOS imaging. 
Once we confidently cleaned their color-magnitude diagrams from field star contamination, 
we estimated their fundamental parameters applying likelihood techniques. We also derived 
their structural parameters from normalized star number density radial profiles. Based on
{\it Gaia} astrometric data, complemented with kinematics information available in the
literature, we computed the 3D components of their space velocities. With similar ages
($\sim$ 2.2 Gyr) and moderately metal-poor overall abundances ([Fe/H] = -1.0 - -0.7 dex),
OGLE-CL-SMC0237 is placed at 2.6 kpc from the SMC center and shares its disk rotation; 
OGLE-CL-SMC0133 is located at 7.6 kpc from the galaxy center and exhibits a kinematics marginally
similar to the SMC rotation disk, while Lindsay~116 placed at 15.7 kpc from the center of
the SMC is facing strong perturbations of its orbital motion with respect to an ordered
rotational trajectory. Furthermore, its internal dynamical evolution would seem to be
accelerated --it seems kinematically older-- in comparison with star clusters in the
outskirts of relatively isolated galaxies. These outcomes lead to conclude that Lindsay~116 
is subject to LMC tides.}

\keywords{technique:photometric -- galaxies: individual: SMC -- galaxies: star clusters}

\titlerunning{Farthest SMC star clusters}  

\authorrunning{A.E. Piatti et al.}           

\maketitle

\markboth{A.E. Piatti et al.:}{Farthest SMC star clusters}

\section{Introduction}           

Star clusters -groups of tens to thousand stars with a common origin in space and time- are 
important constituents of galaxies. By using their astrophysical properties (age, metallicity, 
heliocentric distance, etc), it is possible to trace the formation and evolution of galaxies
where their formed  \citep[see, e.g.,][]{chia2018,oliveiraetal2023}. The former is one of the most active 
field of research in Astrophysics, leading to understand how the Universe formed and evolved. We here pay attention to the Small Magellanic Cloud (SMC), which is with the Large Magellanic Clouds the closest galaxies to the Milky Way, which deserves much of our attention concerning its population of star clusters.

The Small Magellanic Cloud (SMC) is at a heliocentric distance of 62.50 kpc and its depth
is of  $\approx$ 10 kpc \citep{ripepietal2017,muravevaetal2018,graczyketal2020}.
\citet{illescaetal2025} analyzed the public 4m-class imaging SMASH survey database 
\citep{nideveretal2017a} and found three SMC star clusters (OGLE-CL-SMC0133, OGLE-CL-SMC0237 
and Lindsay~116) located along the galaxy line-of-sight with no previous distance estimates. 
Their resulting heliocentric distances place them remarkably outside the SMC at 
21.8 kpc (OGLE-CL-SMC0133), 13.2 kpc (OGLE-CL-SMC0237), and 20.2 kpc (Lindsay~116) from 
the galaxy center, which strikes our understanding about the roles these star clusters have in 
the process of formation and evolution of the galaxy. As can be seen, they are between 2.7 up to 4.4 times farther from the 
SMC center than the outermost known SMC field stars. 

\citet{illescaetal2025} arrived to these results by fitting theoretical isochrones to the 
cluster color-magnitude diagrams (CMDs) -- the deepest existing photometry at the time --, once 
they carefully removed the contamination of field stars by assigning cluster membership 
probability to each measured star by comparison with CMDs of their 
surrounding fields. However, the cluster CMDs show remarkably scattered 
and broad clusters' features (S/N $<$ 15) that led them to estimate heliocentric distances 
($d$) with $\sigma$($d$)/$d$ $\sim$ 0.3. These uncertainties imply that the clusters could be 
part of the SMC main body within 1-2$\sigma$. In order to assess whether the clusters are in 
the SMC or outside it, it is needed $\sigma$($d$)/$d$ $\sim$ 0.05, which in turn implies to 
reach S/N $>$ 50 at the main sequence turnoffs. Attempts to undoubtedly reach S/N $>$ 50 at 
the main sequence turnoff of SMC star clusters similar to those analyzed by 
\citet{illescaetal2025} have been achieved with 8m-class telescopes and high-spatial 
resolution imaging instruments \citep[see, e.g.,][]{mvetal2021}. 

\begin{figure}
\includegraphics[width=\columnwidth]{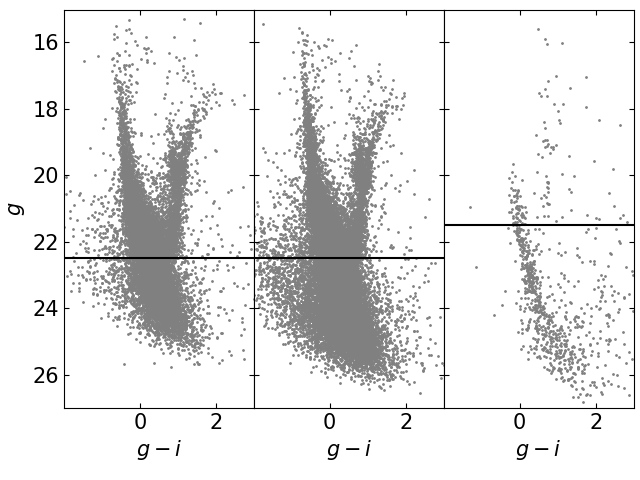}
\caption{Color-magnitude diagrams of the GMOS-S FOV centered on OGLE-CL-SMC0133 (left), 
OGLE-CL-SMC0237 (middle), and Lindsay~116 (right). The solid line represent the limiting magnitude 
reached by the SMASH database in those star cluster fields.}
\label{fig1}
\end{figure}

The confirmation of the derived heliocentric distances of 
these three SMC star clusters, or a percentage of them, or even only one of them, 
opens the possibility to improve our understanding about the formation and evolution of the SMC and its 
dynamical interaction with other galaxies, among others. Star clusters with ages and 
metallicities compatible with being formed in the SMC and now located outside it, are 
witnesses of being stripped from galaxy interaction. 
Finding star clusters beyond the SMC body triggers the speculation that they could have been
stripped from the interaction with the Large Magellanic Cloud (LMC). Indeed, 
\citet{carpinteroetal2013} performed simulations of the dynamical interaction between 
 both Magellanics Clouds 
and their respective star cluster populations. Their models using a wide range of 
parameters for the orbits of both galaxies showed that approximately 15$\%$ of the SMC 
clusters were captured by the LMC. Furthermore, another 20-50$\%$ of its 
star clusters were ejected into the intergalactic medium. More recently, \citet{pl2022}
performed simulations of the orbit of the recently discovered stellar system YMCA-1 and found
that it is a moderately old SMC star cluster that could be stolen by the LMC during any of 
the close interactions between both galaxies, and now is seen to the East from the LMC center.

For the reasons mentioned above, we started an observational campaign with the aim of unveiling whether the 
three analyzed SMC star clusters are indeed far from the SMC. The new heliocentric distances 
derived in this work contribute also to build 
a statistically significant sample of SMC star clusters for a still pending study of
the structure of the cluster population in the SMC \citep{piatti2023c}. 
Likewise, they give new insight into the large-scale 
structure of the area around the SMC.  In Section~2 we describe the collected data 
and different procedures used to obtain the CMDs. Section~3 deals with
the analysis of the star cluster CMDs from the employment of different 
cleaning tools, while in Section~4 we discuss the resulting clusters' heliocentric distances. 
Section~5 summarizes the main conclusions of this work.

\section{Observational data}

We conducted imaging observations with the GEMINI South telescope and the GMOS-S 
instrument \citep[3$\times$1 mosaic of Hamamatsu CCDs; 5.5$\times$5.5 square 
arcmin FOV;][]{allingtonsmithetal2002, hooketal2004, gimenoetal2016} through $g$ and $i$ filters under program GS-2025A-FT-205 (PI: Piatti).
We obtained 8 images per filter and per star cluster in nights with
excellent seeing (0.62$\arcsec$ to 0.97$\arcsec$ FWHM) and photometric 
conditions at a mean airmass of 1.9. We used individual exposure times of 155 sec and 52 sec 
for $g$ and $i$ filters, respectively, to prevent saturation.
The SDSS standard fields E8$\_$a F1, 140000-300000 F2 and PG1633+099 F2 were also
observed alongside bias and sky flats baseline observations.
We reduced the data following the standard procedures (see documentation at the GEMINI Observatory 
webpage\footnote{http://www.gemini.edu} and employed the {\sc gmos/GEMINI} package in 
{\sc iraf/GEMINI}. 
We applied overscan, trimming, bias subtraction, flattening and mosaicing on 
all data images, with previously obtained nightly master calibration images (bias and flats).
Then, we combined (added) all the program images for the same star cluster and filter with 
the aim of gaining in the reachable limiting magnitude. We found that the photometry depth 
increased more than one magnitude with respect to the photometry depth reached by single 
program images.

\begin{figure}
\includegraphics[width=\columnwidth]{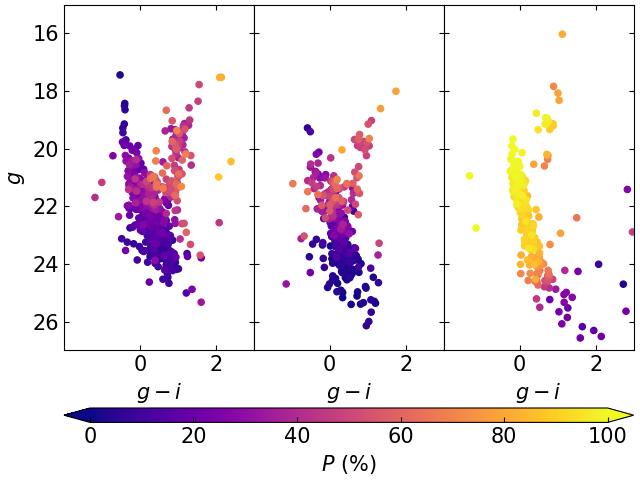}
\caption{Cleaned color-magnitude diagrams of OGLE-CL-SMC0133 (left), OGLE-CL-SMC0237 
(middle), and Lindsay~116 (right). The color bar represents
(in \%) how dissimilar the magnitude and color of a star are from the magnitude and
color distributions of the projected surrounding SMC field-star population.}
\label{fig2}
\end{figure}

The fotometry was obtained as outlined, for example, in \citet{piatti2022c}.
We used routines in the {\sc daophot/allstar} suite of programs \citep{setal90} to find stellar
sources and to fit their brightness profiles with  point-spread-functions (PSFs) in order to
obtain the stellar photometry of each summed image. For each of them, we obtained a preliminary 
PSF derived from the brightest, least contaminated $\sim$ 40 stars, which in turn was used to
derive a quadratically varying  PSF by fitting a larger sample of $\sim$ 100 stars. The two groups 
of PSF stars were selected looking at the image. Then we used the {\sc allstar} program to apply the 
resulting PSF to the identified stellar star clusters. {\sc allstar} generates a subtracted 
image which we employed to find and measure magnitudes of additional fainter stars. We repeated 
this procedure three  times for each summed image. 
Then, we combined all the independent $g,i$ instrumental magnitudes using the stand-alone 
{\sc daomatch} and {\sc daomaster} programs\footnote{Program kindly provided by P.B. Stetson}. 
As a result, we produced one data set per star cluster containing the $x$ and $y$ coordinates 
of each star, the instrumental $g$ and $i$ magnitudes with their respective errors, $\chi$, 
nd sharpness. 
With the aim of removing bad pixels, unresolved double stars, cosmic rays, and background 
galaxies from the photometric catalogs, we kept sources with $|$sharpness$|$ $<$ 0.5. The
photometry of the surviving sources was put into the standard SDSS photometric system
using the following transformation equations:\\

$g$ = 10.55 + $\tilde{g}$, rms = 0.02 mag \\

$i$ = 9.85 + $\tilde{i}$ - 0.05$\times$($\tilde{g}$ - $\tilde{i}$), rms = 0.02 mag.\\

\noindent where $g$,$i$ are the magnitudes in the standard SDSS photometric system, and
$\tilde{g},\tilde{i}$  are the instrumental magnitudes measured by {\sc daophot/allstar} \citep{setal90}.

Figure~\ref{fig1} shows the final photometry for the entire GMOS FOV of the
studied star clusters, which cover a relatively small central circular area of radius 
0.80-1.60 arcmin \citep{bicaetal2020}. As can be seen, the  CMDs resulted 
to be $\sim$ 2.5 up to 4.5 mag deeper than those retrieved by \citet{illescaetal2025}
from the SMASH database. They reveal the presence of a heavy field star contamination 
composed of young to old stars.

\section{Color-magnitude diagram analysis}

The presence of a star cluster in a star field implies the existence of a
local stellar overdensity with distinctive magnitude and color distributions,
whose stars delineate the cluster sequences in the  CMD.
Hence, the  CMD observed along the cluster's line-of-sight 
contains information of both cluster and field stars. This means that if the
CMD features of the star field were subtracted, the
intrinsic characteristics of the cluster CMD would be uncovered.
One possibility of carrying out such a distinction consists in comparing a star field
CMD with that observed along the cluster's line of sight and
then to properly eliminate from the later a number of stars equal to that found 
in the former, taking into account that the magnitudes and colors of the eliminated stars 
in the cluster's CMD must reproduce the respective magnitude and 
color distributions in the star field CMD. The resulting 
decontaminated CMD should depict the intrinsic features
of that cleaned particular field that, in the case of being a star cluster,  are the
well known star cluster's sequences.

\begin{figure}
\includegraphics[width=\columnwidth]{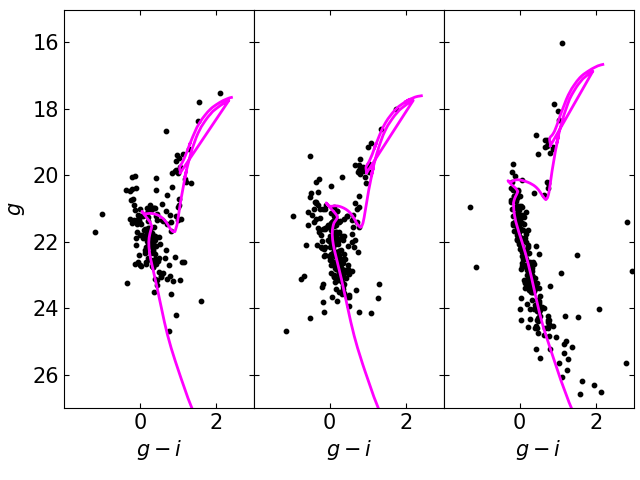}
\caption{Color-magnitude diagrams of OGLE-CL-SMC0133 (left), OGLE-CL-SMC0237 (middle), and Lindsay~116 (right) for stars with $P > 70\%$, 
with the isochrone of \citet[][PARSEC v1.2S]{betal12} corresponding to the best fitted parameters values superimposed  (see text for details).}
\label{fig3}
\end{figure}

We used a circular region around the centers of the studied star clusters with a radius
equals to 0.6$\arcmin$, 0.6$\arcmin$, and 1.0$\arcmin$ for  OGLE-CL-SMC0133, OGLE-CL-SMC0237, 
and Lindsay~116, respectively, which were estimated from previously constructed 
stellar radial profiles based on star counts. 
Similarly, we devised 1000 star field circles with the same radius of the star clusters' circles
and centered at four times the star clusters' radii. They were randomly distributed around the
star clusters' circles. We then built the respective 1000 CMDs with the aim of
considering the variation in the stellar density and in the magnitude and color distributions 
across the star clusters' neighboring regions. The procedure to select stars to subtract from the 
star cluster CMD follows the precepts outlined by \citet{pb12}. 
We applied it using one star field CMD at a time to be compared with the
star cluster  CMD. It consists in defining boxes centered on the magnitude and 
color of each star of the star field  CMD; then it superimposes the boxes 
on the star cluster CMD, and finally it chooses one star per box to 
subtract. In order to guarantee  that at least one star is found within the box boundary, we 
considered boxes with size of ($\Delta$$g$,$\Delta$$(g-i)$) = (1.00 mag, 0.50 mag). 
Whenever. more than one star is found inside a box, the closest one to its center is 
removed. During the choice of the eliminated stars we beard in mind their magnitude 
and color errors by allowing them to have 1000 different values of magnitude and 
color within an interval of $\pm$1$\sigma$, where $\sigma$ represents the errors in 
their magnitude and color, respectively. We also required that the spatial 
positions of the subtracted stars were chosen randomly. In practice, for each field star 
we randomly selected a position in the star cluster's circle and searched for a star to 
subtract within a box of 0.1$\arcmin$ a side. We iterated this loop up to 1000 times, 
if no star was found in the selected spatial box. 

The outcome of the cleaning procedure is a CMD that likely contains only stars that represent the intrinsic features of that
star cluster. From the 1000 different cleaned star cluster  CMDs, we defined
a probability $P$ ($\%$) of being star cluster magnitudes and colors as the ratio $N$/10, where 
$N$ (between 0 and 1000) is the number of times a star was found among the 1000 different 
cleaned  CMDs. Figure~\ref{fig2} shows the resulting cleaned CMDs for the three studied star clusters. They include all the stars
measured within the star clusters' radii and colored according to their respective $P$ values.
We note that the lack of field star contamination reinforces the impression that the star
cluster is actually far away from the SMC.

We used the Automated Stellar Cluster Analysis code \citep[\texttt{ASteCA,}][]{pvp15}
to derive the star cluster fundamental parameters, namely: age, heliocentric distance
and overall metallicity, using stars with membership probabilities $P > 70\%$. 
In order 
to guide \texttt{ASteCA}, we adopted the mean
and dispersion of $E(V-B)$ values from the interpolation in the 
interstellar extinction map built by \citet{sf11}, provided by the NASA/IPAC Infrared Science 
Archive\footnote{https://irsa.ipac.caltech.edu/} for the entire analyzed areas.
\texttt{ASteCA} explores the parameter space  of synthetic  CMDs through 
the minimization of the likelihood function defined by 
\citet[][the Poisson likelihood ratio (eq. 10)]{tremmeletal2013} using a parallel tempering 
Bayesian MCMC algorithm, and the optimal binning \citet{knuth2018}'s method.
\texttt{ASteCA} relies on different compounds to generate the synthetic CMDs. Among
them, it uses the theoretical isochrones computed by
\citet[][PARSEC v1.2S\footnote{http://stev.oapd.inaf.it/cgi-bin/cmd}]{betal12}, the initial 
mass function of \citet{kroupa02} as well as cluster masses in the range 100-5000M$_\odot$, whereas 
binary fractions are allowed in the range 0.0-0.5 with a minimum mass ratio of 0.5.
Table~\ref{tab1} lists the resulting cluster astrophysical properties and the associated
uncertainties, while Figure~\ref{fig3} shows the respective theoretical isochrones 
superimposed onto the star cluster CMDs constructed for stars with 
membership probability $P > 70\%$.

\section{Star cluster properties}

\begin{figure*}
\includegraphics[width=6cm]{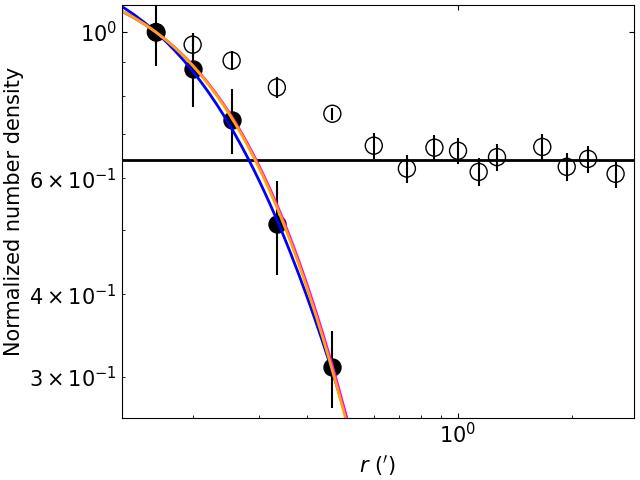}
\includegraphics[width=6cm]{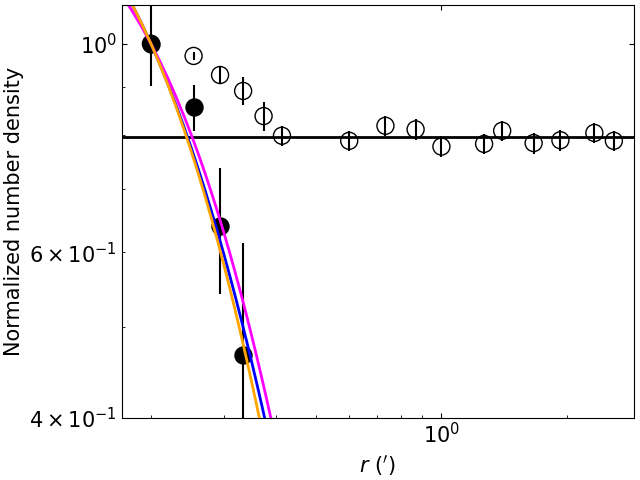}
\includegraphics[width=6cm]{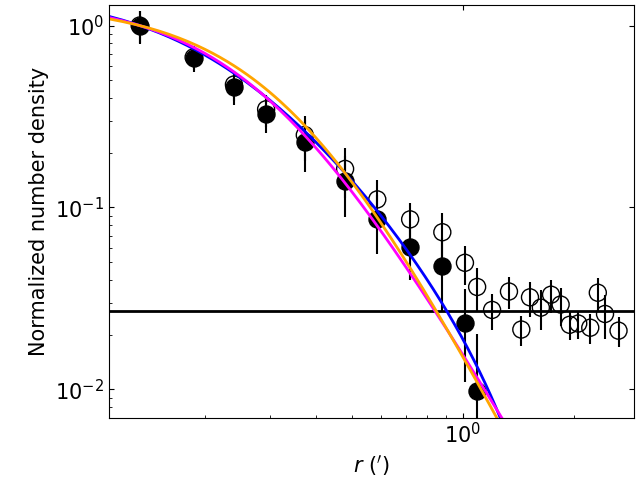}
\caption{Normalized observed and background-subtracted star number density radial profiles of 
 OGLE-CL-SMC0133 (left),  OGLE-CL-SMC0237 (middle) and Lindsay~116 (right)
drawn with open and black filled circles, respectively, with uncertainties represented by error bars. 
The horizontal line represents the adopted mean background level. Blue, orange, and magenta solid 
lines are the best-fitted \citet{king62}'s, \citet{plummer11}'s, and \citet{eff87}'s models, respectively.}
\label{fig4}
\end{figure*}

We took advantage of our deeper GEMINI GMOS images to build
star number density radial profiles for the three studied star clusters.
In order to do that, we counted the number of stars distributed 
inside boxes distributed across the entire observed star cluster fields.
We built several radial profiles, using for each one boxes of a fixed
size, from 0.03$\arcmin$ up to 0.15$\arcmin$ per side, 
increasing in steps of 0.03$\arcmin$ per side. We then averaged all the constructed
radial profiles and estimated their dispersion. 
As expected, the
individual radial profiles built from star counts performed using
smaller boxes resulted smoother toward the innermost clusters' regions, while 
those obtained from larger boxes, traced better the ample clusters' surrounding 
fields. 
Figure~\ref{fig4} depicts the resulting star cluster radial profiles.
We fitted a horizontal line to the background level using points located 
across the whole background region, and adopted the intersection between the observed 
radial profile and the fitted background level as the clusters' radii 
($r_{\rm cls}$, see Table~\ref{tab1}). Extensive artificial star tests were 
previously performed by \citet{petal14} on similar {\sc gmos/GEMINI} data, who found 
that the 50$\%$ completeness level is reached in the innermost regions of
massive LMC star clusters at $g$ $\sim$ 23.5 mag and $i$ $\sim$ 23.8 mag, respectively. 
Hence, we concluded that our radial profiles for the present less populated and less 
crowded star clusters do not suffer from incompleteness.  For the sake of the reader, 
Figure~\ref{fig1_appendix} shows the star counts in $g$ and $i$ for three different
intervals, namely: 1) $r$ $<$ $r_{\rm c}$; 2) $r_{\rm c}$ $<$ $r$ $<$ $r_{\rm cls}$; and
3) $r$ $>$ $r_{\rm cls}$, for  OGLE-CL-SMC0133, OGLE-CL-SMC0237, and Lindsay~116,
respectively.

Finally, we fitted three different
models to the normalized background-subtracted radial profiles, namely: 
the \citet{king62}'s profile model\footnote{
$N \varpropto ({\frac{1}{\sqrt{1+(r/r_{\rm c})^2}} - \frac{1}{\sqrt{1 + (r_{\rm t}/r_{\rm c})^2}}})^2$}, 
which was employed to derive core ($r_{\rm c}$) and
tidal ($r_{\rm t}$) radii;  the \citet{plummer11}'s model helped to derive the
half-light radius ($r_{\rm h}$) from the relation $r_{\rm h}$ $\sim$ 
1.3$\times$a\footnote{$N \varpropto \frac{1}{(1+(r/a)^2)^2}$}; and the 
\citet{eff87}'s model\footnote{$N \varpropto \left(1 + \frac{r^2}{b^2}\right)^{-\gamma/2}$}, 
which  gave  $r_{\rm c}$ $\approx$ $b(2^{2/\gamma} -1)^{1/2}$  and $\gamma$ 
($\gamma$ is the power-law slope at large radii). Table~\ref{tab1} lists the
resulting values of these structural parameters, while Figure~\ref{fig4}
illustrates how they reproduce the star cluster radial profiles.

\begin{table}
\caption{Astrophysical and structural parameters of SMC star clusters.}
\label{tab1}
\small
\begin{tabular}{lccc}
\hline\hline
Parameter         & OGLE-CL- & OGLE-CL- & Lindsay~116 \\
                  & SMC0133  & SMC0237  &              \\
\hline
$E(B-V)^*$ (mag)  & 0.14$\pm$0.01   & 0.11$\pm$0.01   &  0.05$\pm$0.01 \\
$(m-M)_{\rm o}$ (mag)   &  18.70$\pm$0.05 & 18.90$\pm$0.05  & 18.40$\pm$0.05 \\
$d$ (kpc)         &54.95$^{+1.29}_{-1.59}$ & 60.26$^{+1.40}_{-1.37}$ &47.86$^{+1.11}_{-1.09}$ \\
log(age /yr)      & 9.40$\pm$0.05    & 9.35$\pm$0.05  &  9.35$\pm$0.05 \\
$[$Fe/H$]$ (dex)  & -0.70$\pm$0.10  & -0.70$\pm$0.10  & -1.00$\pm$0.10 \\
$r_{\rm c}$ (arcmin)    & 0.30$\pm$0.02   & 0.25$\pm$0.02   & 0.20$\pm$0.02\\
$r_{\rm h}$ (arcmin)    & 0.60$\pm$0.05   & 0.45$\pm$0.02   & 0.45$\pm$0.02\\
$r_{\rm t}$ (arcmin)    & 2.50$\pm$0.20   & 1.50$\pm$0.20   & 2.10$\pm$0.20\\
$r_{\rm cls}$ (arcmin)& 0.65$\pm$0.05   & 0.41$\pm$0.05   & 1.20$\pm$0.05\\
$\gamma$          & 4.00$\pm$0.50   & 4.00$\pm$0.50   & 3.50$\pm$0.50 \\
$c$            &  0.91$\pm$0.03  &  0.77$\pm$0.04   &   1.02$\pm$0.03     \\
$r_{\rm proj}$ ($^{\rm o}$) & 0.88            & 0.82            & 5.84          \\
$r_{\rm deproj}$ (kpc) & 7.6$\pm$0.3    & 2.6$\pm$0.3     & 15.7$\pm$0.2 \\ 
log(M /M$_\odot$) & 3.27            &  3.26           & 2.85    \\
$t_{\rm r}$ (Myr)       & 329$\pm$24      & 242$\pm$9       & 126$\pm$5  \\
age/$t_{\rm r}$   &    7.63$\pm$0.60   &  9.24$\pm$0.65  &  17.75$\pm$1.25   \\
pmra (mas/yr)     & 0.666$\pm$0.071 & 0.818$\pm$0.063 & 1.633$\pm$0.090 \\
pmdec (mas/yr)    & -1.249$\pm$0.030& -1.233$\pm$0.039& -1.098$\pm$0.074\\
\hline
\end{tabular}

\noindent $*$ from NASA/IPAC Infrared Science Archive.

Notes: columns give
interstellar reddening ($E(B-V)$), true distance modulus (($m-M$)$_{\rm o}$),
heliocentric distance ($d$), age (log(age /yr)), overall metallicity ([Fe/H]),
core, half-light, tidal and cluster radii ($r_{\rm c}$,$r_{\rm h}$,$r_{\rm t}$,$r_{\rm cls}$),
EFF gamma value ($\gamma$), concentration parameter ($c$), projected distance ($^{\rm o}$),
deprojected distance ($r_{\rm deproj}$), cluster mass (M/$M_{\odot}$), relaxation
time ($t_{\rm r}$), and proper motions (pmra, pmdec).
\end{table}

We transformed the angular values of $r_{\rm c}$, $r_{\rm h}$, $r_{\rm t}$ and $r_{\rm cls}$ to linear ones
using the expression 2.9 10$^{-4}$$\times$$d$$\times$$r_{\rm c,h,t}$, where $d$ is the star
cluster's heliocentric distance (see Table~\ref{tab1}). We also derived the
star cluster's deprojected distance to the center of the SMC (in kpc) 
using the relation: 

\begin{equation}
r_{\rm deproj} = (d_{\rm smc}^2 +  d^2 - 2 d_{\rm smc} d {\rm cos}(r_{\rm proj}))^{1/2}
\end{equation}

\noindent where $d_{\rm smc}$ is the heliocentric distance of the SMC center
\citep[62.5$\pm$0.8 kpc, ][]{graczyketal2020}.

We computed the star cluster masses (log(M /M$_\odot$)) using the relationships obtained by 
\citet[][equation 4]{metal14} as follows:

\begin{equation}
{\rm log(M /M_{\odot})} = a + b\times {\rm log(age /yr)} - 0.4 (M_{\rm I} - M_{\rm I_{\odot}})
\end{equation}

\noindent where $M_{\rm I}$ is the integrated absolute magnitude in the Johnson $I$ filter  
($M_{\rm I_{\odot}}$ = 4.13 mag), 
and $a$ and $b$ are from Table~2 of \citet{metal14} 
for a representative SMC overall metallicity of $Z$ = 0.004 \citep{pg13}. The absolute
integrated magnitudes were computed from the observed ones as follows:

\begin{equation}
M_{\rm I} = I - A_{\rm I} - 5 {\rm log}(d/10)
\end{equation}

\noindent where $I$ is the observed $I$ integrated magnitude, and $A_{\rm I}$ the
mean interstellar absorption ($A_{\rm I}$ = 1.85$E(B-V)$, see Table~\ref{tab1}). The $I$ values 
were obtained by transforming the measured SDSS $g,i$ integrated magnitudes to the
Johnson $I$ photometric system using theoretical isochrones for the derived star
clusters' ages and metallicities \citep[][PARSEC v1.2]{betal12}. The star cluster
integrated magnitudes were obtained by producing integrated SDSS $g,i$ magnitude growth 
curves from our final photometry (see Section~2). Typical uncertainties turned out to 
be $\sigma$(log(M /$M_{\odot}$)) $\approx$ 0.2.

We also calculated half-light relaxation times using the equation of \citet{sh71}:

\begin{equation}
t_{\rm r} = (8.9\times 10^5 \mathcal{M}_{\rm cls}^{1/2} r_{\rm h^{3/2}}) / (\bar{m} {\rm log}_{10}(0.4\mathcal{M}_{\rm cls}/\bar{m}))
\end{equation}

\noindent where $\bar{m}$ is the average mass of the cluster stars, and units are as follows: 
[$t_{\rm r}$] = Myr; [$r_{\rm h}$] = pc; [masses] = $\msun$. For simplicity we assumed a 
constant average stellar mass of 1.25 M$_{\odot}$, which corresponds to the average between the
smallest and the largest one in the theoretical isochrones of \citet{betal12} representative
of the studied SMC star clusters.
The uncertainties of the computed quantities from eqs. (1) to (4) were obtained by 
performing a thousand
Monte Carlo experiments from which we calculated the standard deviation. Table~\ref{tab1}
list the resulting star cluster masses and relaxation times.

We used the {\it Gaia} DR3 database \citep{gaiaetal2016,gaiaetal2022b} to collect
astrometric information with the aim of deriving the mean star cluster proper motions
in right ascension (pmra) and in declination (pmdec). 
To this respect, we followed the recipes outlined by \citet{piatti2021b}. 
Briefly,
we devised 1000 different adjacent fields as described in Section~2 and used their
Vector Point Diagrams (VPDs) to decontaminate that of the respective star cluster, similarly
as we did to clean the star clusters' CMDs from field stars. From the 1000 different
cleaned star clusters' VPDs we assigned cluster memberships, and kept as cluster
stars those with $P >$ 70$\%$. The mean cluster proper motions and dispersions were
obtained from applying a likelihood approach \citep{pm1993,walkeretal2006}. 
The resulting 
values are listed in Table~\ref{tab1}. Then, we searched the literature for
mean star clusters' radial velocities \citep{parisietal2009,parisietal2015,parisietal2009},
which alongside mean proper motions and heliocentric distances allowed us to
compute the three components of the space velocity vector employing the transformation
equations (9), (13), and (21) in \citet{vdmareletal2002}. We also computed the three
space velocity components of a point placed at the star cluster heliocentric distance 
that rotates according to the SMC rotation disk derived by \citet{piatti2021b}. The
difference between the 3D star cluster velocity components and those of the corresponding
position on the galaxy rotation disk, added in quadrature, has been introduced by 
\citet{piatti2021f} and is called 'residual velocity' index ($\Delta$$V$). We found values of 
$\Delta$$V$ equals to 69.4$\pm$22.0, 25.1$\pm$20.0, and 224.0$\pm$25 km/s, for OGLE-CL-SMC0133, 
0237, and Lindsay~116, respectively.

\section{Discussion}

\begin{figure*}
\includegraphics[width=\columnwidth]{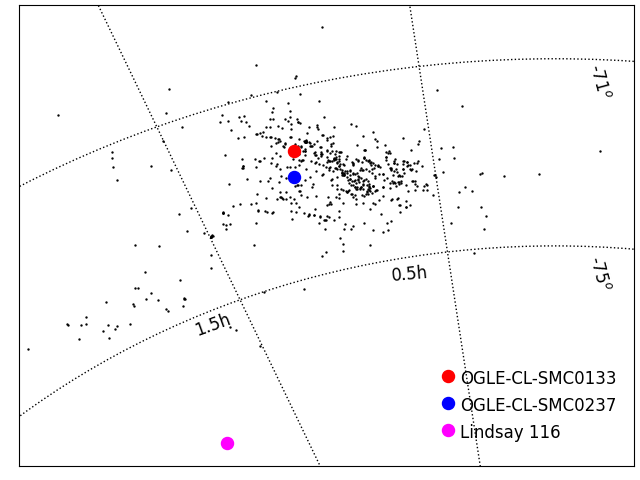}
\includegraphics[width=\columnwidth]{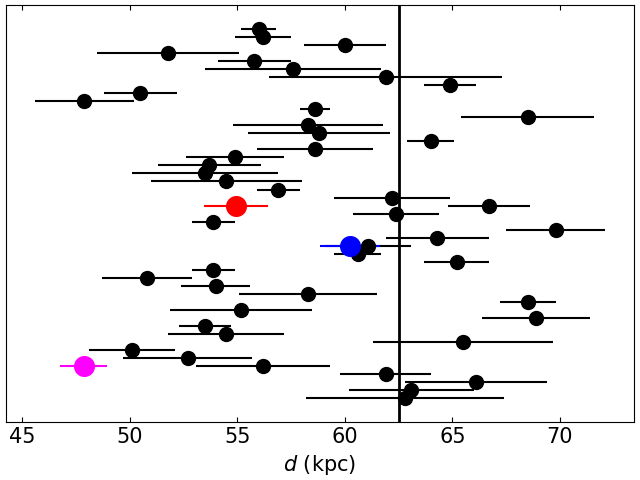}
\caption{Left: Equal-area Hammer projection of the SMC in equatorial coordinates.
Black dots represent the star clusters cataloged in \citet{bicaetal2020}.
Right:
distribution of heliocentric distances of SMC star clusters. The solid vertical line represent
the SMC center derived by \citet{graczyketal2020}.}
\label{fig5}
\end{figure*}

The three studied star clusters are projected onto different SMC composite star fields,
As shown in the left panel of Figure~\ref{fig5},  OGLE-CL-SMC0133 is located toward the
northern edge of the SMC bar, so that a numerous population of young stars is
expected to contaminate its  CMD.  OGLE-CL-SMC0237 is $\sim$ 2.5$\arcmin$
away from the NGC~376's center, a massive young SMC star cluster \citep{miloneetal2023}; while
Lindsay~116 is located in the outermost halo of the SMC. Their  CMDs witness
the aforementioned star contamination (see Figure~\ref{fig2}). 
Particularly, the brighter and low membership probabilities stars ($P$ $<$ 10$\%$) in the 
CMDs of OGLE-CL-SMC0133 and OGLE-CL-SMC0237 likely come from the 
young SMC star
field population and from NGC~376, respectively. Indeed, some residuals from these stars
remain in the cleaned star clusters' CMDs (Figure~\ref{fig3}).
Nevertheless, the main star cluster features are clearly distinguishable.

The derived ages and overall metallicities for the star cluster sample are in very good 
agreement with the values available in the literature and with the recent estimates obtained 
by \citet{illescaetal2025}. Particularly, the difference of their ages and metal content 
and the present ones resulted to be $\Delta$(log(age /yt)) = -0.01 $\pm$ 0.05 and 
$\Delta$[Fe/H] = -0.05 $\pm$ 0.12 dex, respectively. As for the heliocentric distances,
\citet{illescaetal2025} derived distances of 51.1, 75,2 and 42.3 kpc for 
OGLE-CL-SMC0133, 0237 and Linsday~116, respectively. Assuming that the SMC is at a mean
distance of 62.5 kpc \citep{graczyketal2020}, our values (see Table~\ref{tab1}) imply that
the star clusters are closer to the center of SMC. Because the SMC is more extended 
along the line-of-sight and that the line-of-sight depth is $\sim$ 10 kpc, with an average 
depth of 4.3$\pm$1.0 kpc \citep[][and references therein]{ripepietal2017,muravevaetal2018}, 
the new  heliocentric distances put the clusters inside the SMC body.  
Because \citet{illescaetal2025} derived heliocentric distances using 
\texttt{ASteCA}, we discard any difference originated by distinct methodological 
approaches. Instead, we think that they are the result of using shallower and deeper star 
clusters'  CMDs.

We analyzed the distribution of the heliocentric distances ($d$) of  OGLE-CL-SMC0133, 0237, and 
Lindsay~116 using as reference an updated sample of SMC star clusters with distance estimates 
put into an homogeneous scale. For that purpose, we employed as a starting point the 
compilation of star cluster distances of \citet{piatti2023c}, to which we added the results of 
a recent search through the literature carried out by \citet[][their Table~A.1]{illescaetal2025}. 
We finally gathered distance information of 47 SMC star  clusters. The right 
panel of Figure~\ref{fig5} depicts their distance distribution, where we indicated the position 
of the adopted distance of the SMC center with a vertical solid line. As can be seen, the star 
clusters expand along the line-of-sight nearly 20 kpc (from $\sim$ 50 kpc up to 70 kpc), with a 
larger percentage of them distributed throughout the closer galaxy hemisphere.
OGLE-CL-SMC0133 and OGLE-CL-SMC0237 resulted to be located in the main body of the SMC, 
while Lindsay~116 turned out to be one of the closest star clusters to the Sun. 

\begin{figure}
\includegraphics[width=\columnwidth]{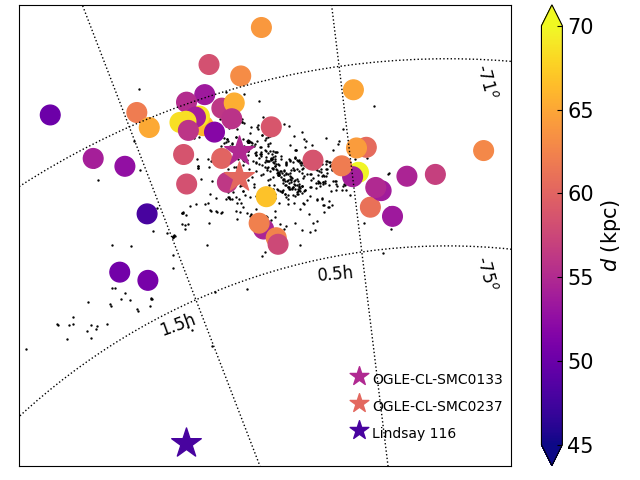}
\caption{Same as left panel of Figure~\ref{fig5}, with colored symbols
indicating star clusters with heliocentric distance estimates. Star symbols represent the studied star clusters.}
\label{fig6}
\end{figure}

Figure~\ref{fig6} shows the spatial distribution of these star clusters as they appear projected 
in  the sky, with their heliocentric distances distinguished by different colored symbols. They 
seem to occupy the SMC body, which makes them valuable in the light of studies 
of the effects of the tidal interaction between both Magellanic Clouds. Indeed, 
Figure~\ref{fig6} suggests that star clusters spanning the whole range of line-of-sight depths
are found throughout the entire SMC main body. However, there are seven star clusters with 
R.A. $>$ 1.5h -- those closest in the sky to the LMC --
that are among the nearest SMC star clusters to the Sun.
In that region there is no studied star clusters farther than $\sim$ 55 kpc from the Sun.
This spatial closeness to the Sun of star clusters located on the Easternmost side of the SMC 
is  in very good agreement with the variation of distances of stars and gas in the
region connecting the SMC and the LMC \citep{ripepietal2017,wagnerkaiser17,omkumaretal2021}.
The LMC is at a mean heliocentric distance of 49.9 kpc \citep{dgetal14}, which implies that
SMC star clusters affected by LMC tides could be found, in terms of heliocentric distances,
closer to the Sun than the SMC.

We probed whether the tidal interaction of the LMC with the SMC has left some 
imprints in the structures and kinematics of the studied SMC star clusters.
At first glance, star clusters under the effects of stronger gravitational fields have
differentially accelerated their internal dynamical evolution, i.e., they appear
dynamically older. This behavior has been observed in Milky Way globular clusters 
\citep{piattietal2019b}, LMC globular clusters \citep{pm2018}, and old SMC star clusters
\citep{piatti2025}, which show their structural/dynamical parameters correlated with the 
distance to the galaxy center or with the tidal force strength. Star clusters toward the 
inner regions of these galaxies are in a more advanced stage of their internal dynamical 
evolution than those located in the outer galaxy regions. Particularly, the structural 
parameter $c$ = log($r_t/r_c$) and the ratio age/$t_r$ seem to decrease with increasing 
distances from the galaxy center. On the other hand, \citet{piatti2021f} 
proposed that the 
residual velocity of SMC star clusters is an index that can be used to 
disentangling whether they are affected by tidal effects caused by the LMC.  He used
the kinematics of 32 SMC star clusters projected toward known tidally perturbed SMC regions
and found that their residual velocities are in general larger than those of SMC clusters
projected toward the SMC main body (see his Figure~3). Indeed, residual velocities larger 
than $\sim$ 60 km/s suggest that star clusters are located in tidally perturbed regions 
or escaping the rotating disk kinematics. This residual velocity cut is in agreement with 
the SMC star cluster dispersion velocity ellipsoid, which has values of 50, 20, and 25 km/s 
along the right ascension, the declination, and the line-of-sight axes, respectively
\citep{piatti2021b}. 

As far as the residual velocity is considered, our resulting values strongly suggest
that the larger the distance of a star cluster to the SMC center, the larger $\Delta$$V$.
OGLE-CL-SMC0237 is located at a deprojected distance of 2.6 kpc and rotates with the
SMC disk. OGLE-CL-SMC0133, at 7.6 kpc from the SMC center, is marginally part of the
rotating disk, while Lindsay~116 is largely confirmed as a star cluster escaping the 
SMC ($r_{deproj}$=15.7 kpc).  We note that these results come from a comparison
of the 3D space velocities of the three clusters with the velocities they would have
if they moved in the SMC rotational disk derived by \citet{piatti2021b} using star
clusters, which as far as we are aware is a self-consistent comparison. Similar analyses
could be attempted by using other derived SMC rotational disks 
\citep[see, e.g.][and references therein]{zivicketal2018,zivicketal2019,zivicketal2021},
not performed here because they are beyond the scope of this work. Because of the position in the sky of Lindsay~116, we 
speculate with the possibility that it is being affected by LMC tides. Moreover,
we examined the structural and dynamics parameter $c$ and age/$t_r$ of the three studied
star clusters and found that Lindsay~116 behaves as it is expected for a star cluster
located in the inner region of the SMC, i.e., it would seem to be in a slightly more advanced
dynamical evolutionary stage. Since Lindsay~116 is one of the outermost SMC star
clusters, we think that the proximity to the LMC could be contributing to accelerate
its internal dynamics. We note that the three studied star clusters have very similar ages
that, in turn, agree very well with that of a star cluster bursting formation event that 
took place  $\sim$ 2.5 Gyr age in both Magellanic Clouds, possibly as a result of a mutual 
close interaction \citep{p11a,p11b,pg13}. During such a bursting episode, the SMC (also 
the LMC) experienced an abrupt increase of its metal content, forming star clusters with
[Fe/H] from $\sim$ 1.2 dex up to -0.6 dex. Nevertheless, according to different simulated
orbits of the LMC/SMC, these galaxies could have experienced some other close encounter 
until the present time \citep[][see Fig.~6, and references therein]{lucchini2024}.

\section{Conclusions}

The SMC is known to have been interacting with the LMC since several Gyr ago \citep{richetal2000,bekkietal2004}.
Recently, \citet{illescaetal2025} derived heliocentric distances of a sample of mostly
unstudied SMC star clusters and found that some of them would be located notably outside
the SMC main body. Because of the interacting nature of this pair of galaxies, the finding of
star clusters orbiting their peripheries could witness the scope of such tidal interactions.
The heliocentric distances estimated by \citet{illescaetal2025} are based on SMASH DR2 data sets,
which imply relatively shallow star cluster CMDs. With the aim of confirming their estimated
star clusters' heliocentric distances, we carried out GEMINI GMOS imaging observations which,
as far as we are aware, resulted in the deepest CMDs of OGLE-CL-SMC0133, 0237, and
Lindsay~116 obtained so far. The 
analysis of these star clusters' CMDs, in addition to structural parameters derived from stellar 
radial profiles built also from the acquired images, and kinematics properties, led us to
conclude as follows:\\

$\bullet$ We confirmed the recently estimated ages and overall metal content of the three 
studied star clusters OGLE-CL-SMC0133, 0237, and Lindsay~116. However, bearing in mind
the 3D extension of the SMC, we found that they are located closer to the SMC center than
previously thought. OGLE-CL-SMC0133 and 0237 are within the SMC main body, while Lindsay~116 
is one of the farthest star cluster of the SMC\\

$\bullet$ When comparing their heliocentric distances with those of an up-to-date sample
of star clusters with distance estimates put into an homogeneous scale, we found that the
three studied star clusters mingle with the other star clusters; Lindsay~116 highlighting
as one of the most external star clusters of the galaxy. Hence, it could have been
subject of tidal stripping from the LMC.\\

$\bullet$ The 3D spatial distribution of the 50 SMC star clusters with heliocentric distance 
estimates reveals that although star clusters are found spanning the whole line-of-sight depth
along any direction toward the SMC, those located in the Easternmost SMC regions (7 in total) 
are systematically closer to the Sun than the SMC center. Such a distance contrast is also 
seen in the spatial distribution of gas and stars connecting the SMC to the LMC.\\

$\bullet$ Based on the behavior found by \citet{piatti2021f} of the residual index
($\Delta$$V$) as a function of the 3D positions of star clusters in the SMC, the presently 
derived $\Delta$$V$ values lead to conclude that the kinematics of OGLE-CL-SMC0237 
resembles that of the rotation of the SMC disk, OGLE-CL-SMC0133 would
be marginally rotating with the SMC disk, while the kinematics of Lindsay~116 would seem to
be largely perturbed as compared to that of a rotational motion.

$\bullet$ Some structural and dynamics properties of Lindsay~116 would also seem to be
reached by tidal effects. Particularly, the concentration parameter $c$ and the age/$t_r$
ratio obtained would point to a star cluster in an internal dynamical evolution stage
compared to those located in the inner galactic regions. However, the loci of Lindsay~116
in the far SMC periphery would imply that an external gravitational potential have 
contributed to make the star cluster kinematically older.

\begin{acknowledgements}
We thank the referee for the thorough reading of the manuscript and
timely suggestions to improve it.

Based on observations obtained at the international GEMINI Observatory, a program of NSF NOIRLab, which is managed by the Association of Universities for Research in Astronomy (AURA) under a cooperative agreement with the U.S. National Science Foundation on behalf of the GEMINI Observatory partnership: the U.S. National Science Foundation (United States), National Research Council (Canada), Agencia Nacional de Investigaci\'{o}n y Desarrollo (Chile), Ministerio de Ciencia, Tecnolog\'{i}a e Innovaci\'{o}n (Argentina), Minist\'{e}rio da Ci\^{e}ncia, Tecnologia, Inova\c{c}\~{o}es e Comunica\c{c}\~{o}es (Brazil), and Korea Astronomy and Space Science Institute (Republic of Korea).

Data for reproducing the figures and analysis in this work will be available upon request
to the first author.

\end{acknowledgements}

\begin{appendix}
\onecolumn
\section{Photometric data properties}
\begin{figure*}[h!]
\includegraphics[width=\textwidth]{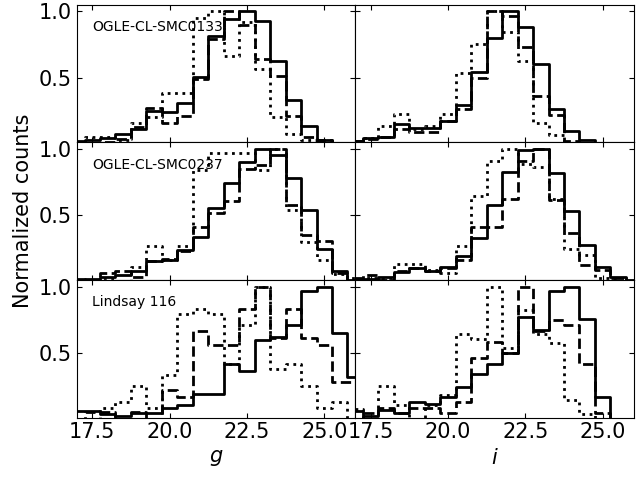}
\caption{Normalized star counts for stars located at $r$ $<$ $r_{\rm c}$ (dotted line), 
$r_{\rm c}$ $<$ $r$ $<$ $r_{\rm cls}$ (dashed line), and $r$ $>$ $r_{\rm cls}$
(solid line), respectively.}
\label{fig1_appendix}
\end{figure*}

Figure~\ref{fig1_appendix} shows the normalized star counts in $g$ and $i$ for
different radial regimes. 

\twocolumn

\end{appendix}

\end{document}